\documentclass[aps, prb, twocolumn, superscriptaddress, floatfix, notitlepage, a4paper, 10pt]{revtex4-2}

\usepackage[utf8]{inputenc}
\usepackage{amssymb}
\usepackage{amsmath}
\usepackage{xcolor}
\usepackage{graphicx}
\usepackage{pdfsync}
\usepackage{hyperref}
\hypersetup{colorlinks, linkcolor={red}, citecolor={blue}, urlcolor={blue}}
\usepackage{bm}
\usepackage{comment}
\usepackage[normalem]{ulem}
\usepackage{mathptmx}

\newcommand{\kk}{{\bm k}}
\newcommand{\rr}{{\bm r}}
\newcommand{\kb}{{k_{\rm B}}}
\newcommand{\effmu}{\mu^{\rm (e)}}
\newcommand{\effeps}{\epsilon^{\rm (e)}}

\begin{document}

\title{Theory of the effective Seebeck coefficient for photoexcited 2D materials: the case of graphene}
\date{\today}

\begin{abstract}
Thermoelectric phenomena in photoexcited graphene have been the topic of several theoretical and experimental studies because of their potential usefulness in optoelectronic applications.
However, available theoretical descriptions of the thermoelectric effect in terms of the Seebeck coefficient do not take into account the role of the photoexcited electron density.
In this work, we adopt the concept of \emph{effective} Seebeck coefficient [G.D. Mahan, \href{https://doi.org/10.1063/1.372988}{J. Appl. Phys. {\bf 87}, 7326 (2000)}] and extend it to the case of a photoexcited two-dimensional (2D) electron gas.
We calculate the effective Seebeck coefficient for photoexcited graphene, we compare it to the commonly used ``phenomenological'' Seebeck coefficient, and we show how it depends on the photoexcited electron density and temperature.
Our results are necessary inputs for any quantitative microscopic theory of thermoelectric effects in graphene and related 2D materials.
\end{abstract}

\author{Andrea Tomadin}
\affiliation{Dipartimento di Fisica, Universit\`a di Pisa, Largo Bruno Pontecorvo 3, I-56127 Pisa, Italy}

\author{Marco Polini}
\affiliation{Dipartimento di Fisica, Universit\`a di Pisa, Largo Bruno Pontecorvo 3, I-56127 Pisa, Italy}
\affiliation{School of Physics \& Astronomy, University of Manchester, Oxford Road, Manchester M13 9PL, United Kingdom}
\affiliation{Istituto Italiano di Tecnologia, Graphene Labs, Via Morego 30, I-16163 Genova, Italy}

\maketitle

\section{Introduction}
\label{sec:introduction}

Thermoelectricity in solid-state systems consists of a group of related effects, where charge currents are coupled to temperature gradients and heat flows.~\cite{goldsmid_book,ziman_book}
In particular, the Seebeck effect consists in the appearance of an electromotive force, measured in terms of a voltage $\Delta V$, at the ends of a junction between two different conductors, if the junction is subjected to heating.

\begin{figure}
(a) \includegraphics[width=0.8\columnwidth]{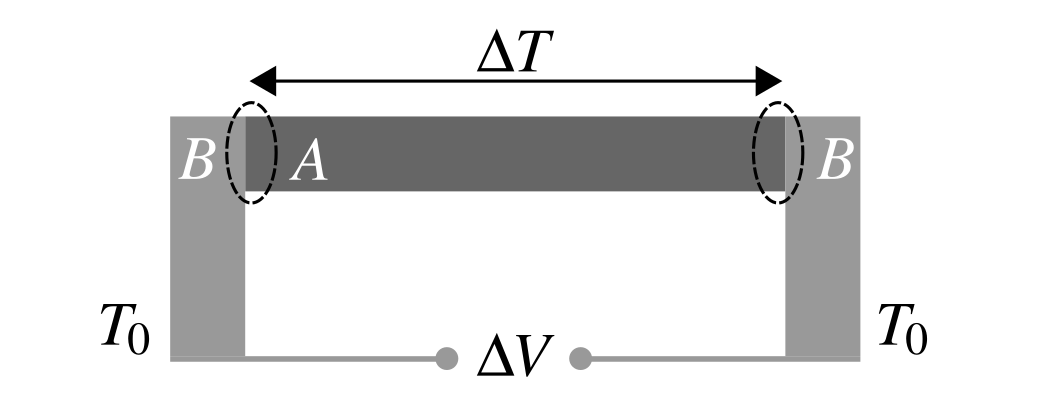} \\
(b) \includegraphics[width=0.8\columnwidth]{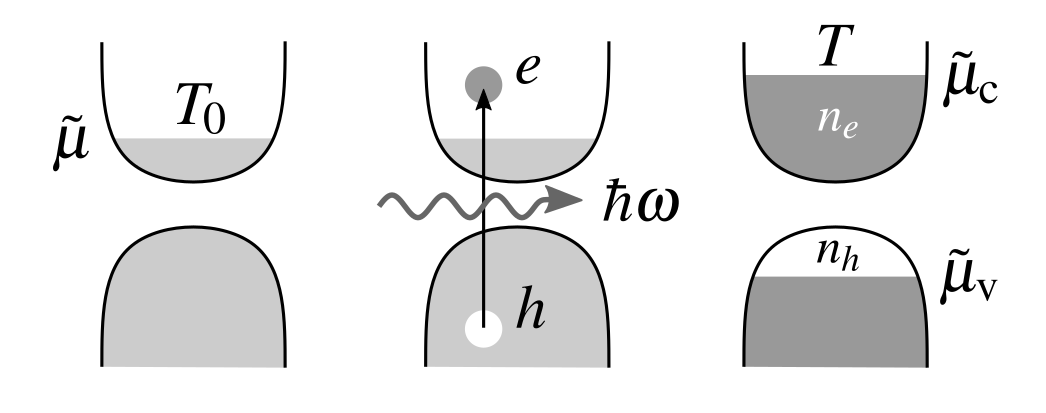} \\
(c) \includegraphics[width=0.8\columnwidth]{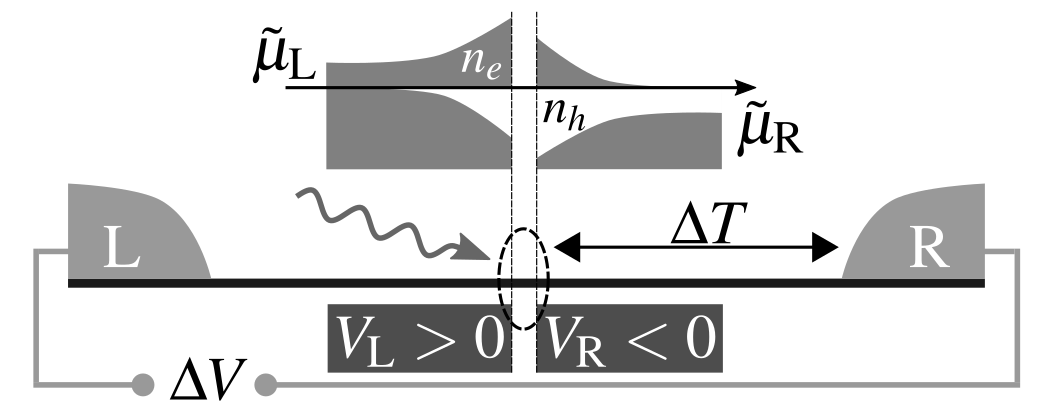}
\caption{\label{fig:setup}
(a) The temperature difference $\Delta T$ between two junctions (dashed ovals) of materials $A$ and $B$ generates the voltage $\Delta V$ between the ends of the leads, maintained at the same temperature $T_{0}$.
(b) Valence and conduction bands in a semiconductor, before, during, and after photoexcitation (from left to right, respectively).
Left: electrons have electrochemical potential $\tilde{\mu}$ and temperature $T_{0}$.
Center: the incoming radiation (wavy line) of angular frequency $\omega$ generates electron-hole ($e$-$h$) pairs.
Right: photoexcited $e$-$h$ pairs relax into a state with higher temperature $T > T_{0}$ and different electrochemical potentials $\tilde{\mu}_{\tau}$ in each band $\tau$.
(c) Radiation impinging onto a p-n junction (dashed oval) in graphene (thick line) induced by a split-gate (dark rectangles) with opposite potentials $V_{\rm L,R}$.
The electron ($n_{e}$, gray) and hole ($n_{h}$, white) densities in space are shown above graphene.
The electrochemical potential ($\tilde{\mu}_{\rm L,R}$) is well-defined away from the junction only.
The voltage $\Delta V$ is measured between the contacts L, R (light gray). }
\end{figure}

The procedure to relate $\Delta V$ to the heating of the junction in a quantitative way is not straightforward.
The standard approach~\cite{goldsmid_book,ziman_book} is to join a wire of material $A$ to \emph{two} leads of material $B$ and consider the temperature difference $\Delta T$ between the two resulting junctions; the voltage $\Delta V$ is measured at the two free ends of the leads, which are held at the same temperature, with the circuit open.
[See Fig.~\ref{fig:setup}(a).]
Then, one finds that $\Delta V = (S_{B} - S_{A}) \Delta T$, where $S_{X}$ is the Seebeck coefficient (or ``thermopower'') of material $X$.
Although only the difference between Seebeck coefficients can be experimentally determined, the expression $\Delta V = - S_{A} \Delta T$ is used if $S_{B}$ is known and its contribution can be subtracted, or is negligible, as in the case of superconducting leads.

The voltage $\Delta V$ is proportional to the difference of the electrochemical potentials $\tilde{\mu} = \mu - e \phi$ at the end of the leads, $\Delta V = - \Delta \tilde{\mu} / e$,~\cite{riess_ssi_1997} where $\mu$ is the chemical potential, $\phi$ the electric potential, and $-e$ the elementary charge.~\cite{fermilevels}
In metallic circuits, continuity of the chemical potential implies that $\Delta V = \Delta \phi$, because the free ends of the leads are kept at the same temperature.~\cite{ziman_book}
However, Mahan and co-workers pointed out~\cite{mahan_jap_2000,cai_prb_2006} that the difference between $\Delta V$ and $\Delta \phi$ cannot be overlooked when evaluating the \emph{local} value of the Seebeck coefficient in a material, and introduced an \emph{effective} (or ``theoretical'') Seebeck coefficient $\bar{S}(\rr)$ defined by the differential relation
\begin{equation}
\nabla_{\rr} \phi(x) \equiv - \bar{S}({\rr}) \nabla_{\rr}T({\rr})~,
\end{equation} 

\clearpage

\vspace{-3em}

\noindent which differs from the local expression of the ``phenomenological'' Seebeck coefficient $S({\rr})$, which is defined by
\begin{equation}
\nabla_{\rr} \tilde{\mu}(x) \equiv e S({\rr}) \nabla_{\rr}T({\rr})~.
\end{equation}
The origin of this difference is that the gradient of the chemical potential, which vanishes when integrated along the junctions' loop, is in general not negligible in a system where the temperature varies in space.

When the materials in the junctions are semiconductors, a more careful consideration of the terms involved in the definition of the Seebeck coefficient might be necessary.
More precisely, heating of one junction can be achieved by shining electromagnetic radiation onto it, locally photoexciting the electrons from the valence to the conduction band.~\cite{sze_book}
(The appearance of the voltage $\Delta V$ is called \emph{photo}-thermoelectric effect, to emphasize its origin.)
In this case, the electronic system is not at equilibrium and the value of the temperature $T$ does not uniquely determine the carrier density in each band.
[See Fig.~\ref{fig:setup}(b).]
In other words, electrons in each band $\tau$ establish a separate electrochemical potential $\tilde{\mu}_{\tau}$ and the standard definition of the Seebeck coefficient outlined above cannot be adopted.
Gurevich and co-workers pointed out this difficulty and used a specific model to calculate the spatial dependence of the electrochemical potentials and hence $\Delta V$.~\cite{gurevich_prb_1995, titov_ijthermo_2016} 
Their analysis is focused on the quasi-neutrality regime, in a one-dimensional geometry, and includes electron-hole recombination processes and boundary effects at the junctions.

While this approach is adequate to a junction's loop, its extension to more complex geometries and current patterns seems cumbersome.
We put forward that a more general calculation framework can be obtained by first evaluating the effective Seebeck coefficient as a function of the system's parameters, and then using this transport coefficient, instead of $S(\rr)$, in the set of transport equations appropriate to the device under consideration.~\cite{jacoboni_book, lundstrom_jeong_book, vasileska_goodnick_book}
Indeed, the effective Seebeck coefficient is naturally well-defined for a photoexcited electron system as long as electrons thermalize at a common temperature $T$.
The time-scale for thermalization due to Coulomb interactions is typically on the order of tens of femtoseconds,~\cite{haug_jauho_book} such that one can consider thermalization to be attained locally and instantaneously with respect to other transport length- and time-scales (e.g.~the transit time through a micron-sized device).
As thermalization with the lattice proceeds through electron-phonon scattering, the electron temperature and the photoexcited electron density evolve in time, and so does the effective Seebeck coefficient.

In this work, we outline the theory of the effective Seebeck coefficient of a photoexcited electron gas and we perform our calculations explicitly in the case of graphene.~\cite{katsnelson_book}
The motivation for focusing on graphene resides in the increasing relevance that this 2D material has been gaining in the field of optoelectronics.~\cite{koppens_naturenanotech_2014, romagnoli_naturereview_2018}
In particular, photodetectors based on the photo-thermoelectric effect can be designed around graphene-based field-effect transistors (GFETs) using a split-gate, that subjects neighboring regions of the graphene channel to different electric potentials, modulating the carrier density and thus creating an effective lateral ``junction'' within graphene.~\cite{gabor_science_2011, cai_naturenanotech_2014, tielrooij_jphys_2015, tielrooij_naturenanotech_2015, brenneis_scirep_2016}
Focusing a laser beam onto the junction leads to the generation of a photo-signal (voltage or current) measured at the contacts on either side (L, R) of the junction.
[See Fig.~\ref{fig:setup}(c).]

The generation of a photo-signal in this setup has been interpreted in terms of the photo-thermoelectric effect, relying on the standard junction picture with different Seebeck coefficients ($S_{\rm L}$, $S_{\rm R}$) in the two graphene regions:~\cite{gabor_science_2011, cai_naturenanotech_2014, tielrooij_jphys_2015, tielrooij_naturenanotech_2015, brenneis_scirep_2016}
(i) a temperature difference $\Delta T$ is established between the illuminated spot and the contacts, which remain at room temperature;
(ii) a potential difference is then generated between each side of the junction and a contact on that side, $\Delta V_{\rm L,R} = -S_{\rm L,R} \Delta T$;
(iii) finally, a voltage $\Delta V = (S_{\rm L} - S_{\rm R}) \Delta T$ proportional to the \emph{difference} of the Seebeck coefficients in the two gated regions is established between source and drain, and a photocurrent flows if the circuit is closed.
Although this picture has been successful in explaining the profile of the photo-signal measured in the experiments, it does not take into account the photoexcited electrons at the junction, which, due to the vanishing density of states of graphene at charge-neutrality, can easily exceed the intrinsic carrier density.
In other words, this approach neglects that the voltage $\Delta V$ is not just a function of the average carrier density induced by the gate on each side of the junction, but also of the laser fluence, the recombination rate, and all the processes that determine how the electrons relax to a single electrochemical potential $\tilde{\mu}_{\rm L,R}$ away from the junction.
For this reason, it is desirable to have a more complete theory of the photo-thermoelectric effect in graphene junctions, which requires the calculation of the effective Seebeck coefficient. The goal of the present work is to provide such theory.

We point out that several works have investigated the electronic properties of photoexcited graphene, including the expression of the dielectric function,~\cite{tomadin_prb_2013} the optical~\cite{svintsov_ape_2014} and terahertz~\cite{tomadin_sciadv_2018} conductivity, and the heat capacity.~\cite{ryzhii_prb_2021}
The expressions discussed in these works are naturally defined in the presence of photoexcitation, by taking into account the band-dependent chemical potentials.
On the contrary, the thermoelectric effect poses a more subtle and fundamental problem, because, as we discussed above, the very definition of the Seebeck coefficient assumes the existence of a single chemical potential.

Our Article is organized as following.
In Section~\ref{sec:theory} we present a general theory of the effective Seebeck coefficient in the presence of photoexcitation.
In Section~\ref{sec:results} we specify our theory to photoexcited graphene and present the results of our calculations for a range of parameters.
Finally, in Section~\ref{sec:conclusions} we draw our main conclusions and identify theoretical and experimental implications of our results.

\section{Theory}
\label{sec:theory}

Let us denote by $f_{\kk\tau}(\rr, t)$ the electron distribution function in a crystal, where $\rr$ is the space coordinate, $t$ the time, $\kk$ the Bloch wave vector in the Brillouin zone, and $\tau$ a multi-index representing all relevant discrete quantum labels, such as band index, spin, and valley.
The distribution function obeys the semiclassical Boltzmann equation,~\cite{ziman_book} which, linearized for a small electric field ${\bm E}(\rr)$ around the quasi-equilibrium solution $\bar{f}_{\kk\tau}(\rr)$, in the absence of a magnetic field, and in the relaxation-time approximation,~\cite{bringuier_eurjphys_2019} reads
\begin{multline}\label{eq:boltzlin}
\partial_{t} \delta f_{\kk\tau}(\rr, t) + {\bm v}_{\kk\tau} \cdot \nabla_{\rr} \delta f_{\kk\tau}(\rr, t) - \frac{e}{\hbar} {\bm E} \cdot \nabla_{\kk} \delta f_{\kk\tau}(\rr, t) \\ 
= - \frac{1}{\tau_{\kk\tau}(\rr)} \delta f_{\kk\tau}(\rr, t)~,
\end{multline}
where $\delta f_{\kk\tau}(\rr, t) \equiv f_{\kk\tau}(\rr, t) - \bar{f}_{\kk\tau}(\rr)$, ${\bm v}_{\kk\tau} \equiv  \hbar^{-1} \nabla_{\kk} \varepsilon_{\kk\tau}$ is the band velocity obtained from the band energy $\varepsilon_{\kk\tau}$, and $\tau_{\kk\tau}(\rr)$ is a relaxation time that depends on the collision processes.~\cite{bringuier_eurjphys_2019}
When electron-electron scattering is the dominant collision term in the semiclassical Boltzmann equation, the quasi-equilibrium solution assumes the Fermi-Dirac form
\begin{equation}\label{eq:fermidirac}
\bar{f}_{\kk\tau}(\rr) = \frac{1}{e^{[\varepsilon_{\kk\tau} - \mu_{\tau}(\rr)]/[\kb T(\rr)]} + 1}~,
\end{equation}
where $\mu_{\tau}(\rr)$ is the chemical potential, $\kb$ the Boltzmann constant, and $T(\rr)$ the temperature.
The quasi-equilibrium distribution function~(\ref{eq:fermidirac}) depends on the band energy but not on $\kk$ explicitly, $f_{\kk\tau}(\rr) = f_{\tau}(\varepsilon_{\kk,\tau}, \rr)$.
(As anticipated in the Introduction, we assume that the temperature is the same in all bands.)

Writing the electric field as ${\bm E}(\rr) = - \nabla_{\rr} \phi(\rr)$, in terms of the electric potential $\phi(\rr)$, and exploiting Eq.~(\ref{eq:fermidirac}), the steady-state solution to Eq.~(\ref{eq:boltzlin}) reads
\begin{multline}\label{eq:distroperturb}
\delta f_{\kk\tau}(\rr) = -\tau_{\kk\tau}(\rr) {\bm v}_{\kk\tau} \cdot \Bigg \lbrace \nabla_{\rr}[\mu_{\tau}(\rr) - e \phi(\rr)] \\
+ \frac{\varepsilon_{\kk\tau} - \mu_{\tau}(\rr)}{\kb T(\rr)} \nabla_{\rr} [\kb T(\rr)] \Bigg \rbrace \left (- \frac{\partial \bar{f}}{\partial \varepsilon_{\kk\tau}} \right )~,
\end{multline}
where $\partial \bar{f} / \partial \varepsilon_{\kk\tau}$ is shorthand for $\left [ \partial \bar{f}_{\tau}(\varepsilon, \rr) / \partial \varepsilon \right ]_{\varepsilon = \varepsilon_{\kk\tau}}$.
The charge current density is given by ${\bm J}(\rr) = \sum_{\tau} {\bm J}_{\tau}(\rr)$, where
\begin{equation}\label{eq:currentdens}
{\bm J}_{\tau}(\rr) = \frac{1}{L^{2}} \sum_{\kk} (-e) {\bm v}_{\kk\tau} \delta f_{\kk\tau}(\rr)
\end{equation}
is the contribution of the electrons belonging to the band $\tau$.
The standard derivation of the transport equations~\cite{ziman_book, jacoboni_book, lundstrom_jeong_book, cantarero_alvarez_2014} proceeds by inserting Eq.~(\ref{eq:distroperturb}) into Eq.~(\ref{eq:currentdens}) and expressing the charge current density in terms of the gradients of the electrochemical potential and the temperature.
We cannot take this step, however, because of the presence of \emph{band-dependent} chemical potentials $\mu_{\tau}$, which cannot be combined with the electric potential $\phi(\rr)$.

To proceed further, let us carefully parametrize $\mu_{\tau}(\rr)$ in terms of the electron density.
Let us denote the electron density in the band $\tau$ by $n^{({\rm i})}_{\tau} + n_{\tau}(\rr) > 0$, where $n^{({\rm i})}_{\tau}$ is the carrier density of the intrinsic system (i.e.~in the absence of electrical or chemical doping, or photoexcitation) at zero temperature, and $n_{\tau}(\rr) \gtrless 0$.
The chemical potential of the band $\tau$ is determined as a function of the density $n_{\tau}(\rr)$ and the temperature $T(\rr)$ by $n^{({\rm i})}_{\tau} + n_{\tau}(\rr) = L^{-2} \sum_{\kk} \bar{f}_{\kk\tau}(\rr)$.
The total carrier density is $n^{({\rm i})} + n(\rr) = \sum_{\tau} [ n^{({\rm i})}_{\tau} + n_{\tau}(\rr) ]$ and the Fermi energy $\varepsilon_{\rm F}(\rr)$ is determined by $n^{({\rm i})} + n(\rr) = L^{-2} \sum_{\kk\tau} \Theta [ \varepsilon_{\rm F}(\rr) - \varepsilon_{\kk\tau} ]$, i.e.~the Fermi energy is the common value of the chemical potentials at zero temperature, which yields the actual total density.
Similarly, we can define a local equilibrium distribution function $f^{(0)}_{\kk\tau}(\rr)$ by substituting $\mu_{\tau}(\rr)$ in Eq.~(\ref{eq:fermidirac}) with a band-independent value $\mu^{(0)}(\rr)$, determined by requiring $n^{({\rm i})} + n(\rr) = L^{-2} \sum_{\kk\tau} f_{\kk\tau}^{(0)}(\rr)$, and, in turn, define $n_{\tau}^{(0)}(\rr)$ by the expression $n^{({\rm i})}_{\tau} + n_{\tau}^{(0)}(\rr) = L^{-2} \sum_{\kk} f_{\kk\tau}^{(0)}(\rr)$.
The \emph{photoexcited electron density} is then
\begin{equation}\label{eq:photoexc_dens}
\delta n_{\tau}(\rr) = n_{\tau}(\rr) - n_{\tau}^{(0)}(\rr)~,
\end{equation}
i.e.~the difference between the actual value of the electron density and the one that \emph{would} be present if the system was at equilibrium.
From the definitions above, it follows that $
\sum_{\tau} \delta n_{\tau}(\rr) = 0$, which represents the fact that photoexcitation does not inject electrons into the system, but promotes them between bands.

Using the equations above, we can implicitly define the band-dependent chemical potential as a function of the Fermi energy, the temperature, and the photoexcited density
\begin{equation}\label{eq:chempotbandfunc}
\mu_{\tau}(\rr) = \mu_{\tau}[\varepsilon_{\rm F}(\rr), T(\rr), \delta n_{\tau}(\rr)]~,
\end{equation}
with the equilibrium, zero-temperature limit given by the Fermi energy, i.e.~$\mu_{\tau}[\varepsilon_{\rm F}(\rr), 0, 0] = \varepsilon_{\rm F}(\rr)$.

Depending on the intrinsic band filling, it might be convenient to perform the calculations in terms of the hole distribution $f_{\kk\tau}^{\rm h}(\rr,t) = 1 - f_{\kk\tau}(\rr,t)$, energy $\varepsilon_{\kk\tau}^{\rm h} = - \varepsilon_{\kk\tau}$, chemical potential $\mu_{\tau}^{\rm h}(\rr) = - \mu_{\tau}(\rr)$, and density $p^{({\rm i})}_{\tau} + p_{\tau}(\rr) = L^{-2} \sum_{\kk} \bar{f}_{\kk\tau}^{\rm h}(\rr)$.
For the photoexcited density it holds that $\delta p_{\tau}(\rr) = -\delta n_{\tau}(\rr)$ and, if a band is entirely filled in the intrinsic system, $p^{({\rm i})}_{\tau} = 0$, it also follows that $p_{\tau}(\rr) = -n_{\tau}(\rr)$.

Using Eq.~(\ref{eq:chempotbandfunc}), we can express the gradient of the chemical potential appearing in Eq.~(\ref{eq:distroperturb}) as
\begin{multline}\label{eq:gradientmu}
\nabla_{\rr} \mu_{\tau}(\rr) = \frac{\partial \mu_{\tau}(\rr)}{\partial \varepsilon_{\rm F}(\rr)} \nabla_{\rr} \varepsilon_{\rm F}(\rr) + \\ \frac{\partial \mu_{\tau}(\rr)}{\partial [\kb T(\rr)]} \nabla_{\rr} [\kb T(\rr)] +
\frac{\partial \mu_{\tau}(\rr)}{\partial \delta n_{\tau}(\rr)} \nabla_{\rr} \delta n_{\tau}(\rr)~.
\end{multline}
In principle, as discussed in the introduction, one would need to solve the complete set of coupled transport equations,~\cite{jacoboni_book, lundstrom_jeong_book, vasileska_goodnick_book} with the charge current density given by Eq.~(\ref{eq:currentdens}) (and a similar expression for the energy current density) to determine the spatial profile of Fermi energy, temperature, and photoexcited density, and calculate the gradients which appear in the right-hand side of Eq.~(\ref{eq:gradientmu}).

In the following, however, we introduce two parametrizations which allow us to focus on the effective Seebeck coefficient only.
First, we assume that the relaxation lengths of the photoexcited density in all bands have the same value, which we parametrize in terms of the relaxation length of the temperature:
\begin{subequations}\label{eq:parametrization}
\begin{equation}
\frac{1}{\delta n_{\tau}(\rr)}\nabla_{\rr} \delta n_{\tau}(\rr) = \alpha \frac{1}{\kb T(\rr)} \nabla_{\rr} [\kb T(\rr)]~,
\end{equation}
where $\alpha$ is a band-independent dimensionless constant.
This is a very reasonable assumption, because the recombination processes which are responsible for the relaxation of the photoexcited carrier density couple different bands, and thus must lead to comparable relaxation lengths.
Second, we assume that the local charge density is neutralized by the gate, adopting the so-called local capacitance approximation~\cite{liu_book}
\begin{equation}\label{eq:gca}
-e n(\rr) = C \phi(\rr), \quad
C = \frac{\epsilon}{4 \pi d}~,
\end{equation}
\end{subequations}
where $d$ and $\epsilon$ are the thickness and the (relative) dielectric constant of the gate layer, respectively, and $C$ is the  capacitance per unit area of the parallel-plate capacitor consisting of the gate and the graphene layer.
(We use Gaussian electromagnetic units.)
With the parametrization~(\ref{eq:parametrization}), the gradient of the chemical potential in Eq.~(\ref{eq:gradientmu}) reads
\begin{multline}\label{eq:grad_chem_pot_exp}
\nabla_{\rr} \mu_{\tau}(\rr) = -\frac{C}{e} \frac{\partial \mu_{\tau}(\rr)}{\partial \varepsilon_{\rm F}(\rr)} \frac{\partial \varepsilon_{\rm F}(\rr)}{\partial n(\rr)} \nabla_{\rr} \phi(\rr) + \\
\left \lbrace \frac{\partial \mu_{\tau}(\rr)}{\partial [\kb T(\rr)]} + \alpha \frac{\delta n_{\tau}(\rr)}{\kb T(\rr)} \frac{\partial \mu_{\tau}(\rr)}{\partial \delta n_{\tau}(\rr)} \right \rbrace \nabla_{\rr} [\kb T(\rr)]~.
\end{multline}

Using Eq.~(\ref{eq:grad_chem_pot_exp}), we can rewrite Eq.~(\ref{eq:distroperturb}) as
\begin{multline}\label{eq:distroperturb2}
\delta f_{\kk\tau} = -\tau_{\kk\tau}(\rr) {\bm v}_{\kk\tau} \cdot \Big\lbrace - \frac{e}{\effeps_{\tau}(\rr)} \nabla_{\rr} \phi(\rr) + \\
\frac{\varepsilon_{\kk\tau} - \effmu_{\tau}(\rr)}{\kb T(\rr)} \nabla_{\rr} [\kb T(\rr)] \Big\rbrace \left (- \frac{\partial \bar{f}}{\partial \varepsilon_{\kk\tau}} \right ) ~,
\end{multline}
where we have defined an effective dielectric constant
\begin{equation}\label{eq:eps_tilde}
\frac{1}{\effeps_{\tau}(\rr)} =  1 + \frac{C}{e^{2}} \frac{\partial \mu_{\tau}(\rr)}{\partial \varepsilon_{\rm F}(\rr)} \frac{\partial \varepsilon_{\rm F}(\rr)}{\partial n(\rr)}
\end{equation}
and chemical potential
\begin{equation}\label{eq:mu_tilde}
\effmu_{\tau}(\rr) = \mu_{\tau}(\rr) - \alpha \frac{\partial \mu_{\tau}(\rr)}{\partial \delta n_{\tau}(\rr)} \delta n_{\tau}(\rr) - \frac{\partial \mu_{\tau}(\rr)}{\partial [\kb T(\rr)]} \kb T(\rr)~.
\end{equation}
We notice that the band dependence has now disappeared from the argument of the gradient operators in Eq.~(\ref{eq:distroperturb2}), so that, inserting Eq.~(\ref{eq:distroperturb2}) into Eq.~(\ref{eq:currentdens}), one finds for the charge current density
\begin{eqnarray}\label{eq:current_lin_eq}
{\bm J}_{\tau}(\rr) =  \bar{\sigma}_{\tau}(\rr) [-\nabla_{\bm r}\phi(\rr)] - \bar{\sigma}_{\tau}(\rr) \bar{S}_{\tau}(\rr) \nabla_{\rr} T(\rr)~.
\end{eqnarray}
The coefficients in Eq.~(\ref{eq:current_lin_eq}) depend on summations over the Brillouin zone.
To make analytical progress, we assume that the relaxation time depends on the band energy, but not on $\kk$ explicitly, $\tau_{\kk\tau}(\rr) = \tau_{\tau}(\varepsilon_{\kk\tau}, \rr)$, and that the band dispersion has azimuthal symmetry, i.e.~$\nabla_{\kk} \varepsilon_{\kk\tau} = (\partial \varepsilon_{k\tau} / \partial k) \nabla_{\bm k} k = \hbar v_{k\tau} \left ( \cos{\theta_{\kk}} \hat{\bm x} + \sin{\theta_{\kk}} \hat{\bm y} \right )$,
where $\theta_{\kk}$ is the polar angle of the wave vector $\kk$.
We obtain the expressions
\begin{subequations}\label{eq:transp_coeff}
\begin{equation}\label{eq:cond_int}
\bar{\sigma}_{\tau}(\rr) = 
\frac{e^{2}}{h} \frac{1}{\effeps_{\tau}(\rr)}
\int_{\infty}^{\infty} d \varepsilon \, 
\frac{\tau_{\tau}(\varepsilon,\rr)}{\hbar} \left (- \frac{\partial f_{\tau}(\varepsilon)}{\partial \varepsilon} \right ) g_{\tau}(\varepsilon)~,
\end{equation}
and
\begin{multline}\label{eq:thpow_int}
\bar{\sigma}_{\tau}(\rr) \bar{S}_{\tau}(\rr) = - \frac{\kb e}{h} \int_{\infty}^{\infty} d \varepsilon \,
\frac{\tau_{\tau}(\varepsilon,\rr)}{\hbar} \frac{\varepsilon - \effmu_{\tau}(\rr)}{\kb T(\rr)} \times \\
\left (- \frac{\partial f_{\tau}(\varepsilon)}{\partial \varepsilon} \right ) g_{\tau}(\varepsilon)~,
\end{multline}
with 
\begin{equation}\label{eq:weight_int}
g_{\tau}(\varepsilon) = \pi \hbar^{2} \frac{1}{L^{2}} \sum_{\kk} v_{k\tau}^{2} \delta(\varepsilon_{k\tau} - \varepsilon)~.
\end{equation}
\end{subequations}
Summing over the band indices, the total (charge) current density reads ${\bm J}(\rr) = \left [ {\textstyle \sum_{\tau}} \bar{\sigma}_{\tau}(\rr) \right ] [-\nabla_{\bm r}\phi(\rr)]
- \left [ {\textstyle \sum_{\tau}} \bar{\sigma}_{\tau}(\rr) \bar{S}_{\tau}(\rr) \right ] \nabla_{\rr} T(\rr)$.
Finally, the open-circuit condition ${\bm J}(\rr) = 0$ leads to the expression for the effective Seebeck coefficient
\begin{equation}\label{eq:eff_see}
\bar{S}(\rr) = \left [ {\textstyle \sum_{\tau}} \bar{\sigma}_{\tau}(\rr) \bar{S}_{\tau}(\rr) \right ] / \left [ {\textstyle \sum_{\tau}} \bar{\sigma}_{\tau}(\rr) \right]
\end{equation}
in terms of Eqs.~(\ref{eq:transp_coeff}).
Before discussing $\bar{S}$ for photoexcited graphene in Sec.~\ref{sec:results}, let us consider two limiting behaviors of Eq.~(\ref{eq:current_lin_eq}).

\begin{figure*}
(a) \includegraphics[width=0.9\columnwidth]{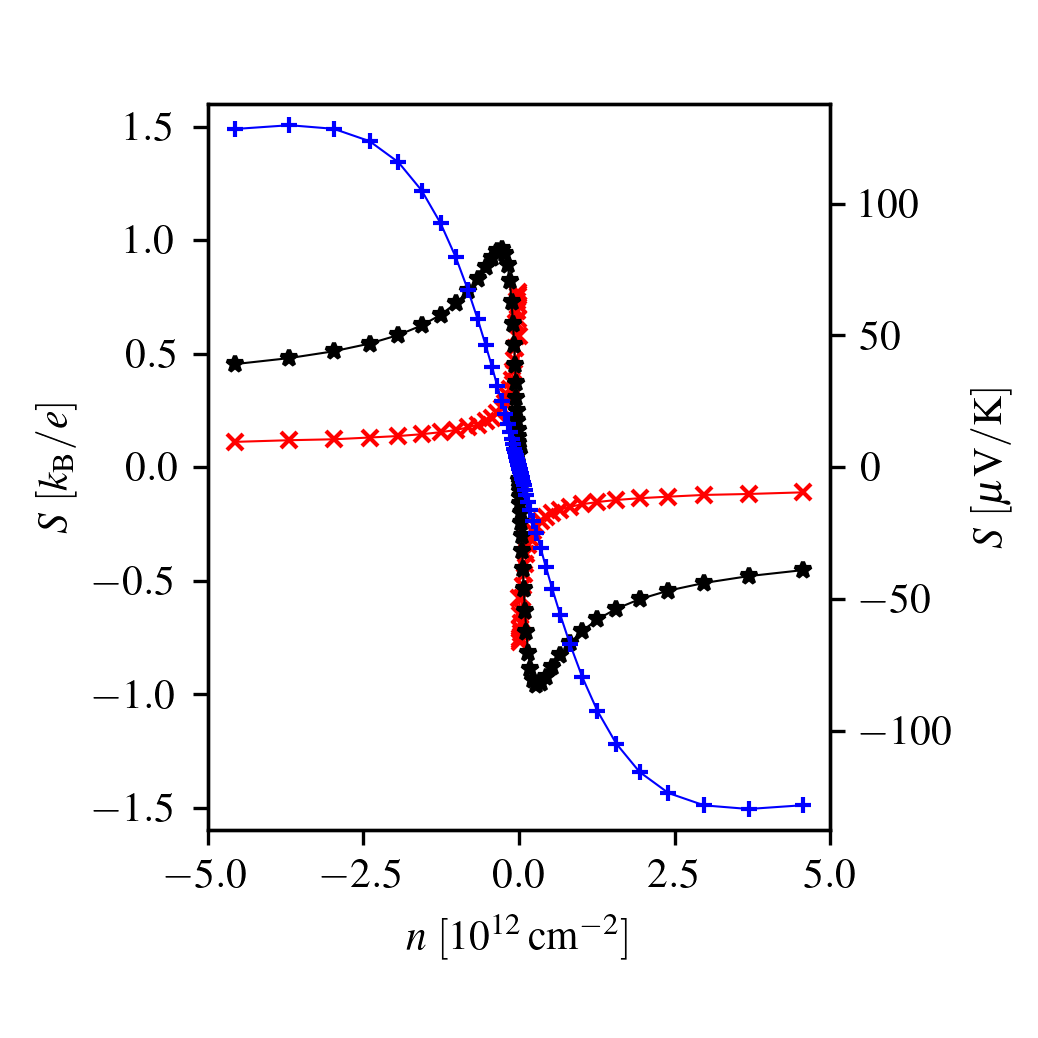}
(b) \includegraphics[width=0.9\columnwidth]{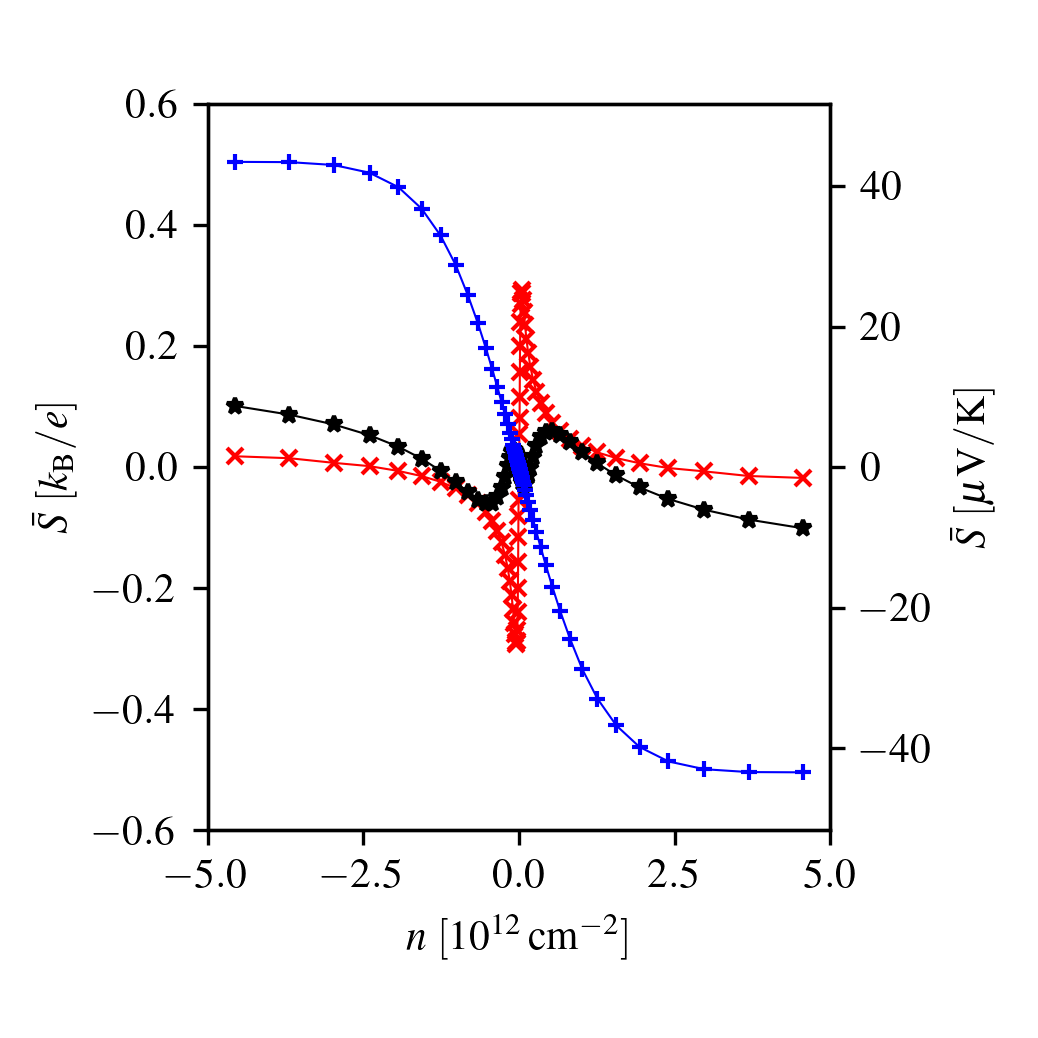}
\caption{\label{fig:equilibrium}
(Color online)
(a) ``Phenomenological'' $S$ and (b) effective $\bar{S}$ Seebeck coefficient as a function of the electron density $n$ at temperature $T = 77~{\rm K}$ (red $\times$), $300~{\rm K}$ (black $\star$), and $1000~{\rm K}$ (blue $+$), for vanishing photoexcited electron density $\delta n_{\rm e} = 0$. }
\end{figure*}

(i) In the local equilibrium limit of vanishing photoexcitation, $\delta n_{\tau}(\rr) = 0$, we have $n_{\tau}(\rr) = n_{\tau}^{(0)}(\rr)$ and $\mu_{\tau}(\rr) =\mu^{(0)}(\rr)$ [defined in the paragraph leading to Eq.~(\ref{eq:photoexc_dens})].
In this limit, a common electrochemical potential $\tilde{\mu}(\rr) = \mu^{(0)}(\rr) - e \phi(\rr)$ for all the bands is well-defined.
The expansion in Eq.~(\ref{eq:gradientmu}) is not needed and, instead of Eq.~(\ref{eq:current_lin_eq}), one finds the standard expression~\cite{ziman_book, jacoboni_book, lundstrom_jeong_book, cantarero_alvarez_2014}
\begin{equation}
{\bm J}_{\tau}(\rr) = \sigma_{\tau}(\rr) \nabla_{\bm r}\tilde{\mu}(\rr) / e - \sigma_{\tau}(\rr) S_{\tau}(\rr) \nabla_{\rr} T(\rr)~,
\end{equation}
with the conductivity $\sigma_{\tau}(\rr)$ and the Seebeck cofficient $S_{\tau}(\rr)$.
The expressions for $\sigma_{\tau}(\rr)$ and $S_{\tau}(\rr)$ are obtained from Eqs.~(\ref{eq:cond_int}) and~(\ref{eq:thpow_int}) with the substitutions $1 / \effeps_{\tau} \to 1$ and $\effmu_{\tau}(\rr) \to \mu^{(0)}(\rr)$.

(ii) If the system is locally at equilibrium \emph{and} the electron density is homogeneous, $\nabla_{\rr }n(\rr) = 0$ and $\nabla_{\rr} \varepsilon_{\rm F}(\rr) = 0$.
In this case, the local-capacitance approximation~(\ref{eq:gca}) leads to $\nabla_{\rr} \phi(\rr) = 0$, which limits the applicability of the theory, unless we also assume $C = 0$ [or a full solution of the Poisson equation is used to relate $n(\rr)$ and $\phi(\rr)$, instead of Eq.~(\ref{eq:gca})].
Under these assumptions, Eq.~(\ref{eq:grad_chem_pot_exp}) simplifies to
\begin{equation}\label{eq:chemgradtemp}
\nabla_{\rr} \mu_{\tau}(\rr) = \frac{\partial \mu_{\tau}(\rr)}{\partial [\kb T(\rr)]} \nabla_{\rr} [\kb T(\rr)]~. 
\end{equation}
This temperature-dependent variation of the chemical potential has been discussed by Mahan and co-workers in Refs.~\cite{mahan_jap_2000, cai_prb_2006}.
Making this dependence explicit allows to rewrite Eq.~(\ref{eq:current_lin_eq}) as
\begin{equation}
{\bm J}_{\tau}(\rr) = \sigma_{\tau}(\rr) {\bm E}(\rr) - \sigma_{\tau}(\rr) \bar{S}_{\tau}(\rr) \nabla_{\rr} T(\rr)~.
\end{equation}
This manipulation proves convenient because the authors of Ref.~\cite{cai_prb_2006} showed that $\bar{S}$ [defined as in Eq.~(\ref{eq:eff_see}) with $\bar{\sigma}_{\tau} \to \sigma_{\tau}$] does not depend on doping and on the material's details.
Moreover, $\bar{S}$ directly relates the temperature gradient to the electric field under the open-circuit condition ${\bm J}(\rr) = 0$.
The consequences of a homogeneous electron density were also investigated in Ref.~\cite{varlamov_epl_2013}, where it was shown that the temperature-dependence of the chemical potential is indeed the crucial function governing the thermoelectric coefficients.

\vspace{-1em}

\section{Results}
\label{sec:results}

To evaluate Eqs.~(\ref{eq:transp_coeff}) for photoexcited graphene, let us first summarize the main properties of the electronic dispersion in graphene:~\cite{katsnelson_book}
the band energy is $\varepsilon_{\kk\lambda} = \lambda \hbar v_{\rm F} \|\kk\|$, where $v_{\rm F}$ is the Fermi velocity and $\lambda$ indicates either the index of the valence (${\rm v}$) and conduction (${\rm c}$) band or the sign $-1$ and $+1$, respectively;
the density of states is $\nu(\varepsilon) = g_{\rm S} g_{\rm V} |\varepsilon| (\hbar v_{\rm F})^{-2} (2 \pi)^{-1}$, where $g_{\rm S} = 2$ and $g_{\rm V} = 2$ are the spin and valley degeneracy, respectively;
we suppose that the spin and valley populations are balanced, hence in Eq.~(\ref{eq:currentdens}) 
$\sum_{\kk\tau} \mapsto g_{\rm S} g_{\rm V} \sum_{\kk\lambda}$;
the Fermi energy as a function of the electron density is $\varepsilon_{\rm F}(\rr) = {\rm sign}[n(\rr)] \hbar v_{\rm F} \sqrt{\pi |n(\rr)|}$;
the intrisic density $n^{({\rm i})}$ vanishes and $n(\rr) \gtrless 0$ corresponds to a zero-temperature electron or hole carrier density, respectively.
Using the expressions above, the quantity in Eq.~(\ref{eq:weight_int}) simplifies to $g_{\lambda}(\varepsilon) = 2 \varepsilon \Theta(\lambda \varepsilon) / (g_{\rm S} g_{\rm V})$, where $\Theta$ is the Heaviside function.

To evaluate Eqs.~(\ref{eq:transp_coeff}), it is also necessary to specify the energy-dependence of the relaxation time $\tau_{\tau}(\varepsilon, \rr)$, which depends on the collision processes.
Following the calculation of the Seebeck coefficient by Das Sarma and co-workers in Ref.~\cite{hwang_rossi_prb_2009}, in this work we focus on electron scattering with Coulomb impurities and we use the expression for the relaxation time in the random-phase approximation, which has been discussed extensively in the literature~\cite{hwang_prl_2007, hwang_prb_2009, dassarma_rmp_2011, kotov_rmp_2012}.
We remark that electron scattering with acoustic and optical phonons in graphene can dominate the transport coefficients in some regimes~\cite{hwang_prb_2008, bistritzer_prb_2009, bistritzer_prl_2009, rana_prb_2009, xie_prb_2016, ponce_repprogphys_2020}.
However, since the focus of the present paper is on the effects of photoexcitation, we defer the investigation of other electron scattering channels to future work.

\begin{figure}[t]
\includegraphics[width=0.9\columnwidth]{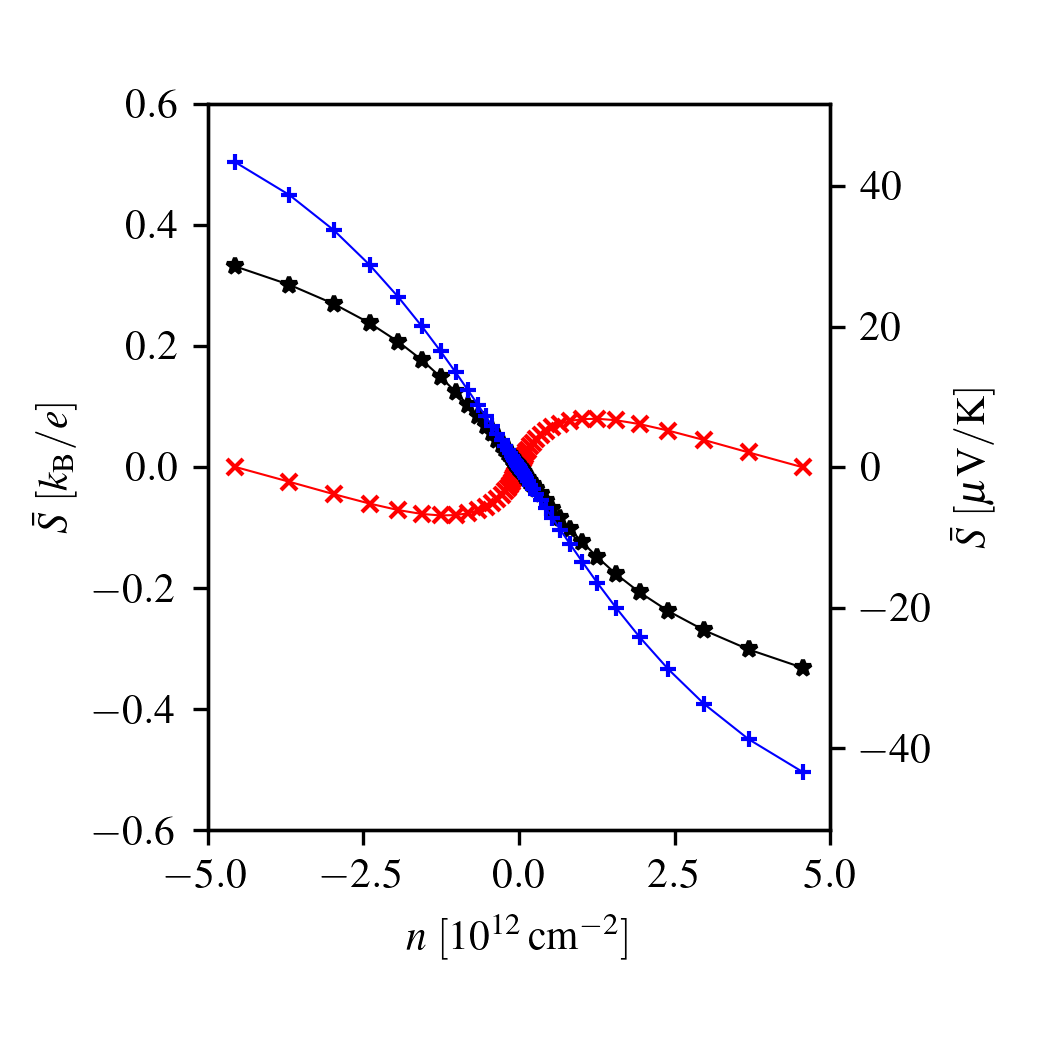}
\caption{\label{fig:temperature}
(Color online)
Effective Seebeck coefficient as a function of the electron density $n$ for a non-equilibrium system with photoexcited electron density $\delta n_{\rm e} = 10^{12}~{\rm cm}^{-2}$ and temperature $T = 300~{\rm K}$ (red $\times$), $1000~{\rm K}$ (black $\star$), and $1500~{\rm K}$ (blue $+$). }
\end{figure}

\begin{figure}
\includegraphics[width=0.9\columnwidth]{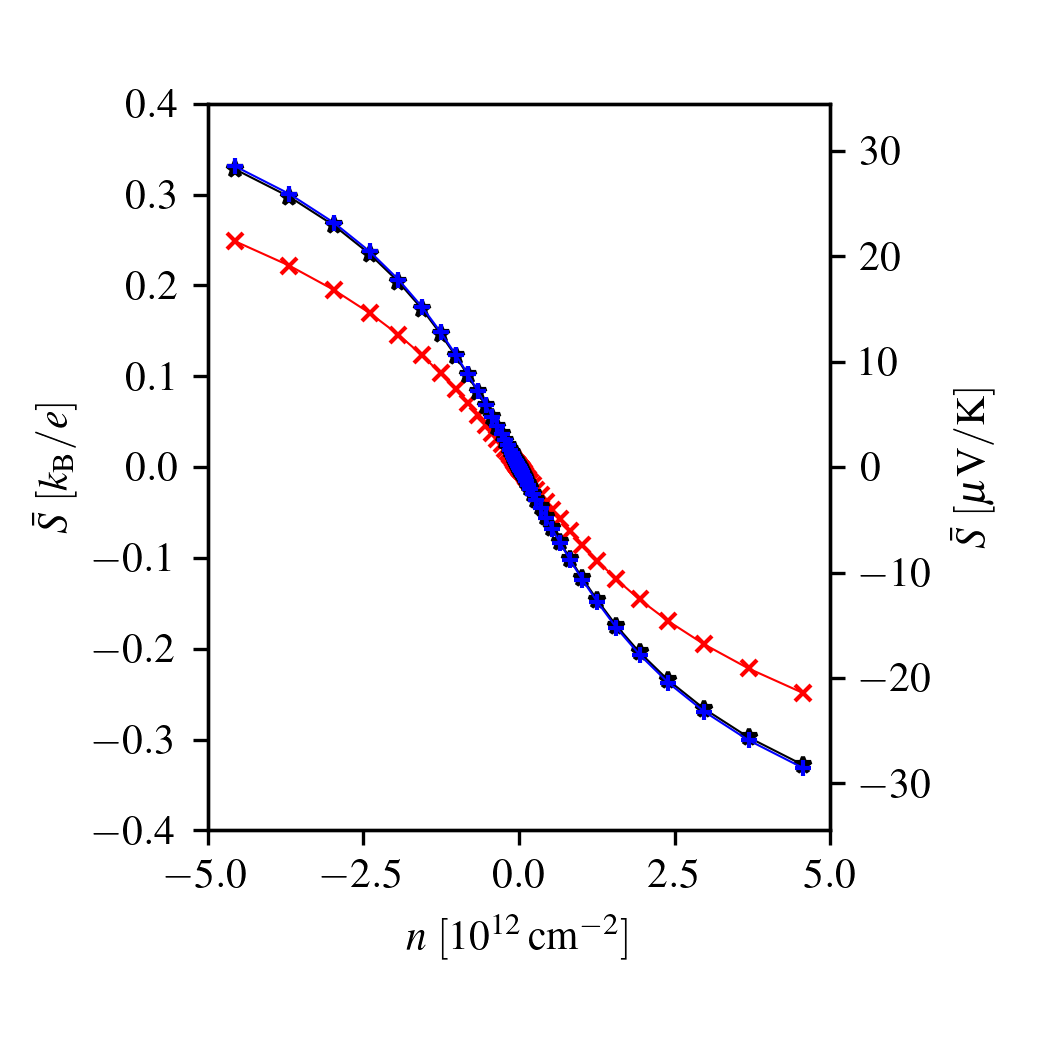}
\caption{\label{fig:gate}
(Color online)
Effective Seebeck coefficient as a function of the electron density $n$ for a non-equilibrium system with photoexcited electron density $\delta n_{\rm e} = 10^{12}~{\rm cm}^{-2}$, temperature $T = 1000~{\rm K}$, and gate layer thickness $d = 1~{\rm nm}$ (red $\times$), $30~{\rm nm}$ (black $\star$), and $300~{\rm nm}$ (blue $+$). }
\end{figure}

\begin{figure}
\includegraphics[width=0.9\columnwidth]{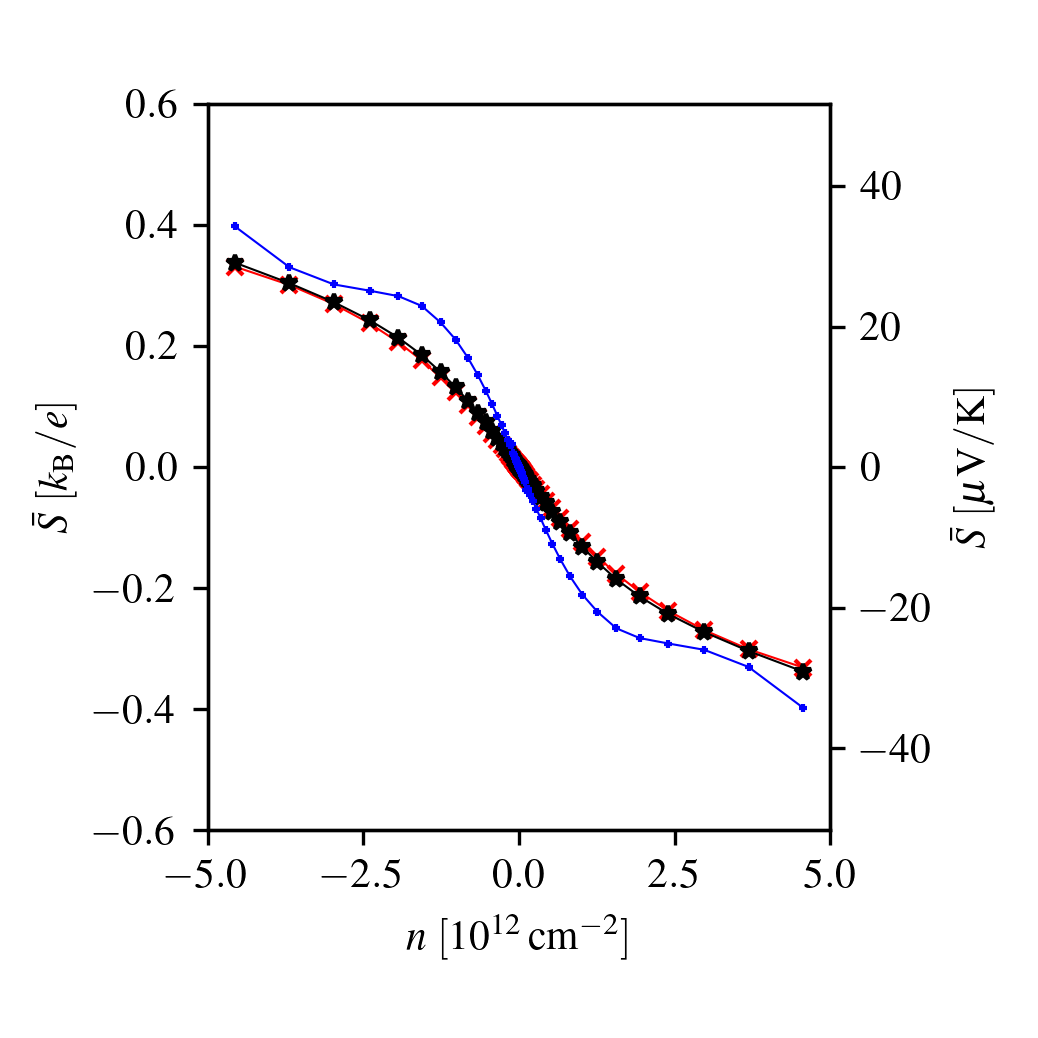}
\caption{\label{fig:relaxation}
(Color online)
Effective Seebeck coefficient as a function of the electron density $n$ for a non-equilibrium system with photoexcited electron density $\delta n_{\rm e} = 10^{12}~{\rm cm}^{-2}$, temperature $T = 1000~{\rm K}$, and relaxation length ratio $\alpha = 0.01$ (red $\times$), $1$ (black $\star$), and $100$ (blue $+$). }
\end{figure}

The results shown in Figs.~\ref{fig:equilibrium}\,--\,\ref{fig:relaxation} have been obtained with the following numerical parameters: Dirac cone slope $\hbar v_{\rm F} = 0.66~{\rm eV}\,{\rm nm}$; dimensionless coupling constant $e^{2} / (\bar{\epsilon} \hbar v_{\rm F}) = 0.8$, where $\bar{\epsilon}$ is the average dielectric constant; Coulomb impurity density $n_{\rm i} = 10^{12}~{\rm cm}^{-2}$; and distance of the Coulomb impurities from the graphene sheet $d_{\rm i} = 1~{\rm nm}$.
In Figs.~\ref{fig:equilibrium}\,--\,\ref{fig:temperature} we take $\alpha = 0$ and $C = 0$ (corresponding to a gate far-removed from the graphene layer), while results with finite $\alpha$ and $C$ are shown in Figs.~\ref{fig:gate} and~\ref{fig:relaxation}, respectively.

\begin{figure}[t]
\includegraphics[width=0.9\columnwidth]{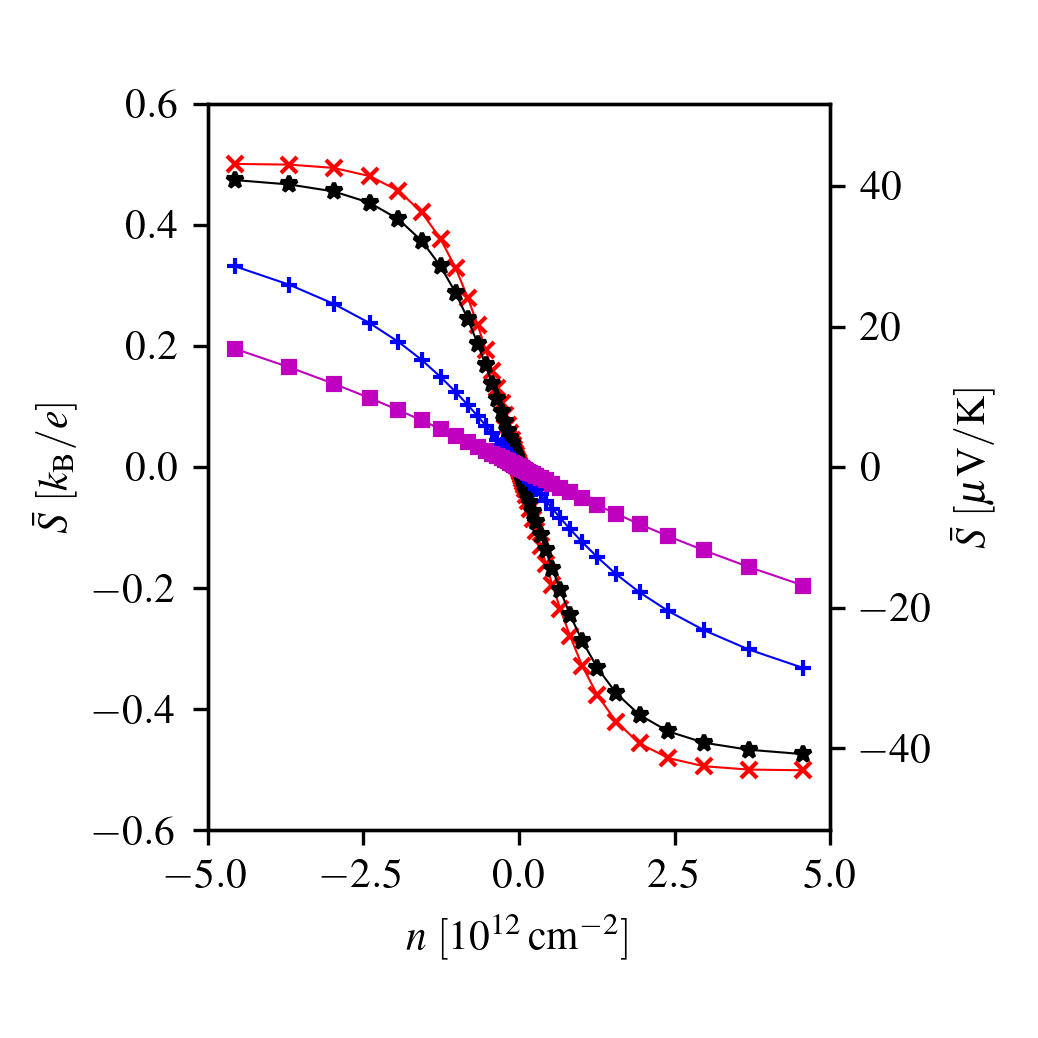}
\caption{\label{fig:photoexc}
(Color online)
Effective Seebeck coefficient as a function of the electron density $n$ for a non-equilibrium system with photoexcited electron density $\delta n_{\rm e} = 10^{10}~{\rm cm}^{-2}$ (red $\times$), $10^{11}~{\rm cm}^{-2}$ (black $\star$), $10^{12}~{\rm cm}^{-2}$ (blue $+$), and $5\times 10^{12}~{\rm cm}^{-2}$ (magenta $\scriptstyle\square$) and temperature $T = 1000~{\rm K}$. }
\end{figure}

Fig.~\ref{fig:equilibrium} compares the ``phenomenological'' and effective Seebeck coefficients at local equilibrium, i.e.~in the absence of photo-excitation.
Let us focus on $n > 0$, as all the profiles are skew-symmetric.
At high temperatures, the two coefficients display a similar dependence on the total density $n$, although $\bar{S}$ is smaller in magnitude.
At low temperatures, however, the behavior of the two coefficients around the neutrality point is the opposite, with $S$ ($\bar{S}$) diverging to negative (positive) values as the density is reduced from larger values, where the two coefficients have the same sign. 
At room temperature, with decreasing $n$, the profile of $\bar{S}$ first follows the low-temperature behavior towards positive values but, approaching the charge neutrality point, bends towards negative values, always remaining much smaller than $S$.
To understand this behavior, let us consider the temperature dependence of the chemical potential for vanishing photoexcitation, which can ben approximated as~\cite{katsnelson_book}
\begin{equation}\label{eq:chempotlim}
\mu_{\rm c}(T) = \left \lbrace \begin{array}{ll}
\varepsilon_{\rm F} \left ( 1 - \frac{\pi^{2}T^{2}}{6 T_{\rm F}^{2}} \right ) & T \lesssim T_{\rm F} \\
\varepsilon_{\rm F} \frac{1}{4 \ln 2} \frac{T_{\rm F}}{T} & T \gtrsim T_{\rm F}~,
\end{array}
\right .
\end{equation}
where $T_{\rm F} = |\varepsilon_{\rm F}| / \kb$ is the Fermi temperature.
Then, in Eq.~(\ref{eq:mu_tilde}) we can substitute
\begin{equation}
\frac{\partial\mu_{\lambda}(\rr)}{\partial [\kb T(\rr)]}  \kb T(\rr) = \left \lbrace \begin{array}{ll}
- \frac{\kb^{2} \pi^{2} T^{2}(\rr)}{3 \varepsilon_{\rm F}(\rr)} & T(\rr) \lesssim T_{\rm F}(\rr) \\
-\mu_{\lambda}(\rr) & T(\rr) \gtrsim T_{\rm F}(\rr)~,
\end{array}
\right .
\end{equation}
which finally yields
\begin{equation}\label{eq:effchemlim}
\effmu_{\lambda}(\rr) = \left \lbrace \begin{array}{ll}
2 \mu_{\lambda}(\rr) & |\varepsilon_{\rm F}(\rr)| \lesssim k_{\rm B} T(\rr) \\
\mu_{\lambda}(\rr) + \frac{\kb^{2} \pi^{2}T^{2}(\rr)}{3 \varepsilon_{\rm F}(\rr)} & |\varepsilon_{\rm F}(\rr)| \gtrsim k_{\rm B} T(\rr)~.
\end{array}
\right .
\end{equation}
Observing how $\effmu_{\lambda}$ enters Eq.~(\ref{eq:thpow_int}), we can rationalize the profiles in Fig.~\ref{fig:equilibrium} as following.
For large density, $\effmu_{\lambda}$ tends to $\mu_{\lambda}$ and the two coefficients $S$, $\bar{S}$ have the same sign.
As the density decreases, $\effmu_{\lambda}$ becomes larger than $\mu_{\lambda}$ in magnitude, contributing a positive (negative) quantity to $\bar{S}$ for $n > 0$ ($n < 0$).
Finally, if the density is sufficiently small, the sign of the integral in Eq.~(\ref{eq:thpow_int}) is determined by the band energy $\varepsilon$ and not the chemical potential.

Fig.~\ref{fig:photoexc} shows the effect of increasing the photoexcited electron density $\delta n_{\rm e}$ on the effective Seebeck coefficient $\bar{S}$.
This information is not available using the ``phenomenological'' Seebeck coefficient, which is not defined unless the electronic system is in local equilibrium.
All profiles correspond to the same electronic temperature $T = 1000~{\rm K}$, which can be easily reached in a photoexcited graphene system.
Indeed, with a heat capacity $c \sim 10^{-6}~{\rm J}/({\rm K}\,{\rm m}^{2})$~\cite{ryzhii_prb_2021}, a temperature increase $\Delta T \sim 10^{3}~{\rm K}$ requires an energy density $\Delta U$ delivered to the electron gas on the order of $\Delta U \sim 10^{-3}~{\rm J}/{\rm m}^{2}$, which is comparable to the product of graphene absorbance $\alpha_{\rm U} \sim 2.3\%$~\cite{katsnelson_book} and a laser pulse fluence ${\cal F} \lesssim 10^{2}~\mu{\rm J}/{\rm cm}^{2}$.
We see that, increasing the photoexcited density, $\bar{S}$ decreases in magnitude.
Of course, decoupling the temperature from the photoexcited density is not possible, in general, in a real experiment, where $\Delta U \sim \hbar \omega \delta n_{\rm e}$, so that $T$ will increase with $\delta n_{\rm e}$.
However, these results show that photoexcitation, although a very effective heating scheme, is not ideal to generate a thermoelectric signal.

Fig.~\ref{fig:temperature} shows the effect of increasing temperature at fixed photoexcited density.
A higher temperature corresponds to a larger thermoelectric signal, which implies a non-linear response of the system to a temperature gradient.
If the temperature of the system was modulated periodically, one could expect higher harmonics of the modulating frequency in the resulting electrical signal.
It is interesting that, at room temperature, the effective Seebeck coefficient changes sign in a large range of densities, being positive (negative) above (below) the charge neutrality point.
This behavior is a signature of the sharp peak shown in Fig.~\ref{fig:equilibrium}(b) and discussed analytically with Eqs.~(\ref{eq:chempotlim})\,--\,(\ref{eq:effchemlim}) in the local equilibrium case.
The solution of the transport equations in specific geometries might be necessary to appreciate the consequences of this change of sign.
Indeed, if the temperature decreases in space (away from the laser spot where photoexcitation takes place) on a shorter length-scale than the photoexcited density, (i.e.~$\alpha \ll 1$,) it can happen that neighboring spatial regions are characterized by effective Seebeck coefficients with opposite sign, leading to unexpected thermoelectric current patterns.

Finally, Figs.~\ref{fig:gate} and~\ref{fig:relaxation} show the dependence of $\bar{S}$ on the two coefficients $\alpha$ and $C$ introduced in Eq.~(\ref{eq:parametrization}) to parametrize the gradient expansion~(\ref{eq:grad_chem_pot_exp}) of the chemical potential.
We see that $\bar{S}$ changes very weakly when $\alpha$ and $C$ vary over several orders of magnitude.
This reassures us that the results shown in Figs.~\ref{fig:equilibrium}\,--\,\ref{fig:temperature}, where we fixed $\alpha = 0$ and $C = 0$, are generally valid.
Moreover, the weak dependence of the results on $\alpha$ and $C$ emphasizes the role of the temperature-dependence of the chemical potential, Eq.~(\ref{eq:chemgradtemp}), which was discussed in Refs.~\cite{mahan_jap_2000, cai_prb_2006, varlamov_epl_2013} in the absence of photoexcitation.

\vspace{-1.5em}

\section{Conclusions and Perspectives}
\label{sec:conclusions}

In this Article we have formulated a theory of the Seebeck effect when the temperature gradient is generated by photoexcitation.
We have discussed how the standard definition of the Seebeck coefficient is not adequate in this case, because it assumes the existence of a well-defined chemical potential, which is missing in a photoexcited multi-band electron system.
We have thus formulated our theory in terms of the \emph{effective} Seebeck coefficient $\bar{S}$, first introduced by Mahan and co-workers,~\cite{mahan_jap_2000} which is properly defined for photoexcited electron systems, and we have provided explicit results in the case of graphene.
We have shown that $\bar{S}$ decreases at fixed temperature with increasing photoexcited density, implying that photoexcitation, although effective at increasing the electronic temperature, is not ideal to generate a thermoelectric signal.
Moreover, we have found that $\bar{S}$ displays a sign change at lower temperatures, which could lead to unexpected thermoelectric current profiles in specific geometries.
To investigate this issue, it will be necessary to use $\bar{S}$ in the framework of a complete set of transport equations.

Recently, an experimental technique has been introduced which allows to measure the Seebeck effect locally, without resorting to photoexcitation,~\cite{harzheim_nanolett_2018, harzheim_2dmater_2020} demonstrating that the Seebeck coefficient can undergo variations of several orders of magnitude due to purely geometric constraints on the electronic motion.
This technique is based on a scanning Joule heating element and is characterized by high spatial resolution, comparable to the length-scale of a junction produced by a split-gate.~\cite{gabor_science_2011, cai_naturenanotech_2014, tielrooij_jphys_2015, tielrooij_naturenanotech_2015, brenneis_scirep_2016}
Moreover, local and ultra-fast measurement of the electronic temperature has also been recently demonstrated.~\cite{aamir_nanolett_2021}
Hence, it is becoming experimentally feasible to measure the thermoelectric signal generated by heating the \emph{same} spot on a graphene sample, in the presence or absence of a concurrent photoexcited density, extracting valuable information on the local electronic relaxation processes.
The theory described in this work is the first one that can consistently treat both cases on equal footing.

\vspace{-1.5em}

\begin{acknowledgments}
Useful discussions with K.-J. Tielrooij are gratefully acknowledged.
This work was supported by the European Union's Horizon 2020 research and innovation programme under grant agreement no.~881603 -- GrapheneCore3.
\end{acknowledgments}

\end{document}